\documentclass[onecolumn,superscriptaddress]{revtex4}
\usepackage[dvips]{graphicx}
\usepackage{amsmath,amsthm,amssymb}
\usepackage[english]{babel}
\usepackage{wrapfig}
\usepackage{hyperref}

\begin{document}

\title[Perturbed pendulum-like motions of a rigid body]
{PERTURBED PENDULUM-LIKE MOTIONS \\ OF A RIGID BODY ABOUT A FIXED POINT}
\author{Igor N. Gashenenko}
\email{gashenenko@iamm.ac.donetsk.ua} \affiliation{Institute of Applied Mathematics and Mechanics of National Academy of Sciences of
Ukraine, Luxemburg Str. 74, Donetsk 83114, Ukraine}

\begin{abstract}
This paper is devoted to a detailed investigation of the perturbed pendulum-like motions of a heavy rigid body about a fixed point.
Canonical variables that allow one to simplify the analysis of homoclinic and heteroclinic orbits are introduced. Characteristic
pro\-perties of perturbed pendulum-like motions of the body in inertial space are studied. A qualitative description of asymptotics
of pendulum-like motions in a neighbourhood of split separatrices is given.
\end{abstract}
\date{\today}


\maketitle

\tableofcontents

\section{Introduction}\label{sec1}

Differential equations of motion, referred to the principal axes inertia of the body at its fixed point,
have the following form
\begin{equation}
\label{eq} \frac{d{\bf M}}{d t}={\bf M}\times{\boldsymbol \omega}+\mu({\boldsymbol
\gamma}\times {\bf r}),\,\,\,\,\,\,\frac{d{\boldsymbol  \gamma}}{d t}={\boldsymbol
\gamma}\times{\boldsymbol  \omega},
\end{equation}
where $\,{\boldsymbol  \omega}$ is the angular velocity, ${\bf M}=A{\boldsymbol  \omega}$
is the angular momentum of the rotated body, $\,A={\rm
diag}(A_1,A_2,A_3)$ is the tensor of inertia, ${\boldsymbol  \gamma}$ is the unit vector directed vertically upward, ${\bf r}$ is the unit vector to the body's center of gravity, $\mu$ is a product of the weight and the distance from a fixed point to the gravity center.

Let us start the investigation of a body motion, whose mass distribution is restricted by the following conditions
\begin{equation}
\label{cond_P}A_1>A_3>A_2,\,\,\,r_1\geq0,\,\,\,r_2\geq0,\,\,\,
r_3=0,\,\,\,\mu\ll 1.
\end{equation}
If the conditions (\ref{cond_P})  have been fulfilled, and the following equalities $ \omega_1=\omega_2=\gamma_3=0$ are used as the initial values of the variables,  so the body is rotated about the fixed horizontal axis as a physical pendulum:
\begin{equation}\label{5:3}
\begin{split}
&M_1=M_2=\gamma_3=0,\,\,\,\,\,\,M_3=\pm\sqrt{2A_3(h_0+\mu)}\,\,\text{dn}\tau,\\
&\gamma_1=r_1[2\,\text{sn}^2\tau-1]\mp2\,r_2\,\text{sn}\tau\,\text{cn}\tau,\,\,\,\,\,\gamma_2=r_2[2\,\text{sn}^2\tau-1]\pm2\,r_1\,\text{sn}\tau\,\text{cn}\tau,\\
&\tau=(t-t_0)\sqrt{(h_0+\mu)/(2A_3)},\,\,\,\,\,k=\sqrt{2\mu/(h_0+\mu)},
\end{split}
\end{equation}
where the energy constant  $h_0>\mu$ is defined by the initial conditions.

For the further investigation of a rigid body motions in a neighborhood of the integrable Euler case the canonical Andoyer-Deprit variables are successfully used~\cite{Dep,Koz80,Zig80}. We will use the following notations:  $G$ is the modulus of the angular momentum $\,\bf{M},$
$L$ is a projection of the angular momentum $\,\bf{M}\,$  on a moving axis $\,Oz,$ $H$ is a projection of the vector  $\,\bf{M}\,$ on an upward-directed vertical;
the variables canonically conjugate to $\,G, L, H\,$ are the angles $\,g, l, h\,$ in range $[0,\,2\pi).$ The dependency between phase variables and Andoyer-Deprit's ones are expressed in the following formulas:
\begin{equation}
\label{depri_G}
M_1=\sqrt{G^2-L^2}\sin{l},\,\,\,\,\,M_2=\sqrt{G^2-L^2}\cos{l},\,\,\,\,\,M_3=L,
\end{equation}
\begin{equation}
\label{depri_g}
 \begin{split}
&\gamma_1=(\sin{q}\cos{p}+\sin{p}\cos{q}\cos{g})\sin{l}+\cos{q}\sin{g}\cos{l},\\
&\gamma_2=(\sin{q}\cos{p}+\sin{p}\cos{q}\cos{g})\cos{l}-\cos{q}\sin{g}\sin{l},\\
&\gamma_3=\,\sin{q}\sin{p}-\cos{p}\cos{q}\cos{g},
 \end{split}
\end{equation}
where $\,\,\sin{p}=L/G,\,\,\,\sin{q}=H/G.$ The inverse transformation is given by
\begin{equation}
\label{gamma_dep}
 \begin{split}
&G=\vert{\bf M}\vert,\,\,\,g=\arcsin\left({\frac{(M_2\gamma_1-M_1\gamma_2)}{\sqrt{M_1^2+M_2^2}}\frac{\vert\bf{M}\vert}{\vert\bf{M}\times{\boldsymbol \gamma}\vert}}\right),\\
&L=M_3,\,\,\,l=\textrm{arctan}(M_1/M_2),\,\,\,H=\bf{M}\cdot{\boldsymbol \gamma}.
 \end{split}
\end{equation}
We introduce new parameters $\,\alpha_0,\beta_0,\gamma_0$
that characterize the mass distribution:
\begin{equation}
\label{phi}
 \begin{split}
&\alpha_0=2\,\textrm{arctan}\left(\frac{\sqrt{A_1(A_3-A_2)}}{\sqrt{A_2(A_1-A_3)}}\right),\,\,\,\,\beta_0=2\,\textrm{arctan}\left(\frac{r_2}{r_1}\right)\in[0,\pi],\\
&\gamma_0=2\,\textrm{arctan}\left(\frac{\sqrt{A_3-A_2}}{\sqrt{A_1-A_3}}\right),\qquad\quad\text{where}\,\,\,\,\,\pi>\alpha_0>\gamma_0>0.
 \end{split}
\end{equation}
Immediately from (\ref{phi}) the following relations can be found
$$\displaystyle A_1=A_3\,\frac{\sin^2{\displaystyle\frac{\alpha_0}{2}}}{\sin^2{\displaystyle\frac{\gamma_0}{2}}},\,\,\,A_2=A_3\,\frac{\cos^2{\displaystyle\frac{\alpha_0}{2}}}
{\cos^2{\displaystyle\frac{\gamma_0}{2}}},\,\,\,r_1=\cos{\frac{\beta_0}{2}},\,\,\,r_2=\sin{\frac{\beta_0}{2}}.$$

If $\alpha_0=\beta_0,$ then the rigid body mass distribution is subjected to the Hess conditions (the center of gravity belongs to the perpendicular, drawn from a fixed point to the circular section of gyration ellipsoid). If  $\gamma_0=\beta_0,$ then body mass distribution is subjected to the Grioli conditions (the center of gravity belongs to the perpendicular, drawn from a fixed point to the circular section of inertia ellipsoid). Moreover, from the triangle inequality $A_2+A_3>A_1$  we determine an additional restriction for the parameters:
$$\cos{\frac{\alpha_0}{2}}>\cos^2{\frac{\gamma_0}{2}}. $$\vskip-2mm

With regard to (\ref{depri_G}), (\ref{depri_g}) the Hamiltonian of the mechanical system will have the following form
\begin{equation}
\label{h_d}
 \begin{split}
\mathcal{H}_*=&\frac{1}{2}\left[(G^2-L^2)(A_1^{-1}\sin^2{l}+A_2^{-1}\cos^2{l})+A_3^{-1}L^2\right]+\\
+&\mu\left[\sin{q}\cos{p}\sin{(l+\beta_0/2)}+\cos{q}\sin{p}\sin{(l+\beta_0/2)}\cos{g}+\right.\\
+&\left.\cos{q}\cos{(l+\beta_0/2)}\sin{g}\right].
 \end{split}
\end{equation}
We changed the sequence of the principal inertia axes in moving basis $\,Oxyz,$  therefore the present canonical variables $\,G, L, H,$ $ g, l, h\,$  are distinct from the standard variables, used in the research works~\cite{Dep,Koz80,Zig80,BM,G_G_S81}. For rotations of a rigid body about its intermediate inertia axis, the canonical variables (\ref{gamma_dep}) are degenerated. Later, it will be found, that such variables simplify the study of the perturbed pendulum motions.
Under conditions (\ref{cond_P}) the rotations (\ref{5:3}), as it is known, are unstable. Due to this, the investigation of qualitative properties of the perturbed body motions in the fixed space is of our special interest and also determines the main objective of this paper.

\section{New canonical variables}\label{sec2}
Using the generating function
$$
S_2=(2l+\beta_0)J_1+(g-l-\beta_0/2)J_2+h J_3
$$
we shall find the relations
\begin{equation}
\label{canon}
 \begin{split}
&\;L=\frac{\partial S_2}{\partial l}\equiv
2J_1-J_2,\,\,\,\,\,G=\frac{\partial S_2}{\partial g}\equiv
J_2,\,\,\,\,\,H=\frac{\partial S_2}{\partial h}\equiv J_3,\\
\theta_1&=\frac{\partial S_2}{\partial J_1}\equiv
2l+\beta_0,\,\,\,\theta_2=\frac{\partial S_2}{\partial J_2}\equiv
g-l-\beta_0/2,\,\,\,\theta_3=\frac{\partial S_2}{\partial J_3}\equiv h,\\
 \end{split}
\end{equation}
which define the canonical transformation to the variables
$\,(J_i,\theta_i),\,i=\overline{1,3}$. So from (\ref{canon}) the following expressions are obtained
$$
l=\frac{\theta_1-\beta_0}{2},\,\,\,g=\frac{1}{2}\,\theta_1+\theta_2,\,\,\,h=\theta_3,\,\,\,J_1=\frac{G+L}{2},\,\,\,J_2=G,\,\,\,J_3=H.
$$
Let us introduce the dimensionless variables
$J_i^\prime=J_i\sqrt{(A_1-A_2)/(A_1A_2)},\,i=\overline{1,3},$ and the notation for time $\,t^\prime=t\sqrt{(A_1-A_2)/(A_1A_2)}.$
In order to reduce the record, the primes of dimensionless variables should be omitted.
In the new variables the Hamiltonian (\ref{h_d}) is transformed to the following form
\begin{equation}
\label{h_new}
 \begin{split}
\widetilde{\mathcal{H}}=&J_1(J_2-J_1)\left[\cos{(\theta_1-\beta_0)}-\cos{\alpha_0}\right]+\frac{1}{2}\,\varkappa J_2^2+\\
+&\mu\left[J_1\sin{(\theta_1+\theta_2)}+(J_2-J_1)\sin{\theta_2}
\right]\frac{\sqrt{J_2^2-J_3^2}}{J_2^2}\,+\\
+&2\mu\,\frac{J_3\sqrt{J_1(J_2-J_1)}}{J_2^2}\,\sin{\frac{\theta_1}{2}},
 \end{split}
\end{equation}
where
$$\,\varkappa=\frac{A_1A_2}{A_3(A_1-A_2)}\equiv\frac{\sin^2{\alpha_0}}{2\,(\cos{\gamma_0}-\cos{\alpha_0})}>0.$$
At $\,J_3=0\,$ level the Hamiltonian (\ref{h_new}) will be expressed as follows
\begin{equation}
\label{h0_new}
\begin{split}
\mathcal{H} =\, &\frac{\varkappa}{2}J_2^2+
J_1(J_2-J_1)\left[\cos{(\theta_1-\beta_0)}-\cos{\alpha_0}\right]+\\
+
&\frac{\mu}{J_2}\left[J_1\sin{(\theta_1+\theta_2)}+(J_2-J_1)\sin{\theta_2}
\right],
 \end{split}
\end{equation}
We will suppose that angle  $\,\theta_{2}\,$ is changed in range $[0,2\pi),$ and the alteration of
variables $\,J_{1,2}\,$  is restricted by the following  inequalities  $J_2>0,\,J_2\geq J_1\geq 0.$
At $J_3=0$ level, the differential equations of rigid body motion are the Hamilton equations
$$
\dot{\theta_i}=\frac{\partial\mathcal{H}}{\partial
J_i},\,\,\,\dot{J_i}=-\frac{\partial\mathcal{H}}{\partial
\theta_i},\,\,\,i=1,2,
$$
we obtain them in the explicit form:
\begin{equation}
\label{eq1}
\begin{split}
&\dot{\theta_1}=(J_2-2J_1)\left[\cos{(\theta_1-\beta_0)}-\cos{\alpha_0}\right]+
\frac{\mu}{J_2}\left[\sin{(\theta_1+\theta_2)}-\sin{\theta_2}\right],\\
&\dot{J_1}=J_1(J_2-J_1)\sin{(\theta_1-\beta_0)}-\mu\frac{J_1}{J_2}\cos{(\theta_1+\theta_2)},\\
&\dot{\theta_2}=\varkappa J_2+
J_1\left[\cos{(\theta_1-\beta_0)}-\cos{\alpha_0}\right]-
\mu\frac{J_1}{J_2^2}\left[\sin{(\theta_1+\theta_2)}-\sin{\theta_2}\right],\\
&\dot{J_2}=-\frac{\mu}{J_2}\left[J_1\cos{(\theta_1+\theta_2)}+(J_2-J_1)\cos{\theta_2}\right].
 \end{split}
\end{equation}

If we know the system (\ref{eq1}) solution, so the cyclic coordinate $\,\theta_3\,$ can be easily defined by the quadrature from the equation
\begin{equation}
\label{th3}
\dot{\theta_3}=2\mu\frac{\sqrt{J_1(J_2-J_1)}}{J_2^2}\sin{\frac{\theta_1}{2}}.
\end{equation}

\section{Particular solutions of the Hamiltonian system (\ref{eq1})}\label{sec3}
For the small values of the parameter $\,\mu,$  the angular momentum modulus $\,\vert {\bf M}\vert=J_2(t)\,$  remains in a small neighborhood of its initial value $\,J_2^0=J_2(t_0).$
Let us consider the exact solutions of the system (\ref{eq1}), which correspond to the energy levels, close to $\,h_0=\varkappa J_2^2(t_0)/2>\mu.$

\subsection{Unperturbed system solutions} 
\begin{figure}[ht]\centerline{
\includegraphics[scale=1.85,clip,viewport=15 26 105
95,keepaspectratio]{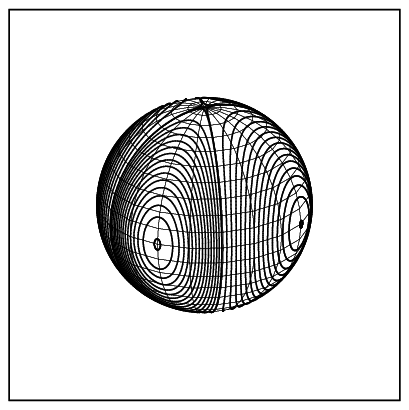}\hskip5mm
\includegraphics[scale=1.65,clip]{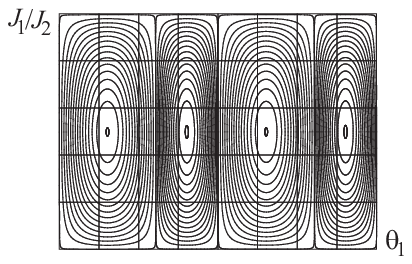}
}{\footnotesize {\it a}}\hskip70mm{\footnotesize {\it
b}}\caption{Phase orbits of the Euler case: \textit{a)}
at the sphere $\vert\bf{M}\vert^2=\textrm{const}$,\quad \textit{b)} at the plane $\,\mathbb{R}^2(\theta_1,J_1/J_2)$.}\label{p1}
\end{figure}

As $\,\mu=0\,$ we have the unperturbed system of differential equations
\begin{equation}
\label{eq1_0}
\begin{split}
&\dot{\theta_1}=(J_2-2J_1)\left[\cos{(\theta_1-\beta_0)}-\cos{\alpha_0}\right],\\
&\dot{\theta_2}=\varkappa J_2+J_1\left[\cos{(\theta_1-\beta_0)}-\cos{\alpha_0}\right],\\
&\dot{J_1}=J_1(J_2-J_1)\sin{(\theta_1-\beta_0)},\,\,\,\,\,\dot{J_2}=0.
 \end{split}
\end{equation}
This system  is completely integrable: it describes the rigid body motion in the integrable Euler case. The general solution of the system (\ref{eq1_0}) may be expressed in terms of theta functions of time. The unperturbed separatrices, used below, correspond to doubly-asymptotic solutions of the system (\ref{eq1_0}):
\begin{equation}
  \label{sol_1}\begin{split}
1)\,\,&\theta_1=\beta_0-\alpha_0,\,\,\,
J_1=J_{2}^0/\left(e^{u(t-t_0)+v}+1\right),\qquad\\
&\theta_2=\varkappa J_{2}^0(t-t_0)+\theta_{2}^0,\,\,\, J_2=J_{2}^0;
\end{split}
\end{equation}
\begin{equation}
  \label{sol_2}
  \begin{split}
2)\,\,&\theta_1=\alpha_0+\beta_0,\,\,\,
J_1=J_{2}^0/\left(e^{-u(t-t_0)+v}+1\right),\qquad\\
&\theta_2=\varkappa J_{2}^0(t-t_0)+\theta_{2}^0,\,\,\,
J_2=J_{2}^0;\end{split}
\end{equation}
\begin{equation}
  \label{sol_12}\begin{split}
3)\,\,&\theta_1=\beta_0-\alpha_0+2\pi,\,\,\,
J_1=J_{2}^0/\left(e^{u(t-t_0)+v}+1\right),\,\,\\
&\theta_2=\varkappa J_{2}^0(t-t_0)+\theta_{2}^0,\,\,\, J_2=J_{2}^0;
\end{split}
\end{equation}
\begin{equation}
  \label{sol_22}
  \begin{split}
4)\,\,&\theta_1=\alpha_0+\beta_0+2\pi,\,\,\,
J_1=J_{2}^0/\left(e^{-u(t-t_0)+v}+1\right),\\
&\theta_2=\varkappa J_{2}^0(t-t_0)+\theta_{2}^0,
J_2=J_{2}^0,\end{split}
\end{equation}
where $\,\theta_i^0,J_i^0\,$ are the initial values of the variables at $t=t_0,$  the constant parameters are denoted through
$$u=J_2^0\sin{\alpha_0}>0,\,\,\,\,\,v=\ln{\frac{(J_2^0-J_1^0)}{J_1^0}}\in (-\infty,\infty).$$
In the limiting cases from the relations (\ref{sol_1})--(\ref{sol_22}) the following equalities can be found:
$$
1) \;\; \theta_1=\beta_0-\alpha_0,\,\,\,J_1=0; \,\,\qquad\qquad 1^*)
\;\; \theta_1=\beta_0-\alpha_0,\,\,\,J_1=J_2;\qquad\,\,
$$
$$
2) \;\; \theta_1=\beta_0+\alpha_0,\,\,\,J_1=0; \,\,\qquad\qquad 2^*)
\;\; \theta_1=\beta_0+\alpha_0,\,\,\,J_1=J_2;\qquad\,\,
$$
$$
3) \;\; \theta_1=\beta_0-\alpha_0+2\pi,\,\,\,J_1=0; \qquad 3^*) \;\;
\theta_1=\beta_0-\alpha_0+2\pi,\,\,\,J_1=J_2;
$$
$$
4) \;\; \theta_1=\beta_0+\alpha_0+2\pi,\,\,\,J_1=0; \qquad 4^*) \;\;
\theta_1=\beta_0+\alpha_0+2\pi,\,\,\,J_1=J_2.
$$
At Fig.~\ref{p1} the phase orbits of unperturbed system of equations (\ref{eq1_0}) at the sphere $\vert{\bf{M}}\vert^2=\textrm{const}$  and at the plane $\,\mathbb{R}^2(\theta_1,J_1/J_2)$ for the values $\,A_1=2,\,A_2=1,\,\,A_3=1.5\,$ can be reviewed. The separatrix, composed of two intersecting circles, Fig.~\ref{p1},\,{\it a}, is transformed in the vertical segments, Fig.~\ref{p1},\,{\it b}.

\subsection{Physical pendulum} Pendulum-like rotations of a rigid body are expressed explicitly by Jacobi elliptic functions of time:
\begin{equation}
  \label{sol_3}
  \begin{split}
1)\,\,&J_1=0,\,\,\,J_2=\sqrt{2\,(h_0+\mu)/\varkappa}\,\,\text{dn}\tau,\,\,\,\sin{\theta_{2}}=2\text{sn}^2\tau-1,\\
2)\,\,&J_1=J_2,\,\,\,J_2=\sqrt{2\,(h_0+\mu)/\varkappa}\,\,\text{dn}\tau,\,\,\,\sin{(\theta_{1}+\theta_{2})}=2\text{sn}^2\tau-1,
\end{split}
\end{equation}
where
$\,\tau=(t-t_0)\sqrt{(h_0+\mu)\varkappa/2},\,\,\,\,\,k=\sqrt{2\mu/(h_0+\mu)}.$

From the relations (\ref{depri_G}),(\ref{canon}) it follows, that the angle $\,\theta_1\,$ hasn't been defined for the equalities $\,J_1=0$ and $\,J_1=J_2.$  This degeneracy is not essential, as far as the body's position and velocity, relevant to the mentioned equalities, are uniquely defined by the smaller numbers of the phase variables.

If $\,J_1=o(\mu),$  then by using the integral (\ref{h0_new}) we exclude the value of $\,J_2\,$ from the 1st and 3rd equations (\ref{eq1}). Due to the variables transformation $\,y=\textrm{ctg}(\theta_1/2),\,\tau=\theta_2\,$, we can obtain Riccati differential equation
\begin{equation}\label{eq_Ric1}
y^{\prime}+a_0y^2+a_1(\tau)y+a_2(\tau)=0,\end{equation}
\begin{equation}\nonumber
  \begin{split}
&a_0=\frac{1}{2\varkappa}(\cos{\beta_0}-\cos{\alpha_0}),\,\,\,\,\,a_1(\tau)=\frac{1}{\varkappa}\sin{\beta_0}+\frac{\mu\cos{\tau}}{2\,(h-\mu\sin{\tau})},\\
&a_2(\tau)=-\frac{1}{2\varkappa}(\cos{\alpha_0}+\cos{\beta_0})-\frac{\mu\sin{\tau}}{2\,(h-\mu\sin{\tau})}.
\end{split}
\end{equation}

If $\,J_2-J_1=o(\mu),$  then the similar Riccati equation (\ref{eq_Ric1}) will be obtained after transformation $\,y=\textrm{ctg}(\theta_1/2),$ $\tau=\theta_1+\theta_2.$ The dependence $\,y(\tau)\,$ allows investigating $\,\theta_1(\theta_2)\,$  in the arbitrarily small neighborhood of pendulum motions. Also we can note that  an analytical study of the variational equations for pendulum motions of a rigid body, and the simplest cases analysis of its integrability are presented in the paper~\cite{Dok68}.

Small perturbations of the exact solutions, considered in the cases {\it
a),b)}, lead to the formation of asymptotic solutions at the energy levels, close to $\,h_0=\varkappa J_2^2(t_0)/2>\mu.$ Let us introduce formulas describing the asymptotic solutions of the equations (\ref{eq1}) to the first-order approximation.

\subsection{The solutions, close to pendulum type}  We get the approximation formulas for small values $\,\mu.$  Let us put $\,\theta_2^*(t)=\varkappa J_{2}^0(t-t_0)+\theta_{2}^0.$ The periodic solutions of the system (\ref{eq1}) are the following:
\begin{equation}
\label{tt}
\begin{split}
&\widetilde{T}_1:\,\,\,\,\,
\theta_1=\beta_0+\varepsilon\alpha_0+\mu(d_{11}\sin{\theta_2^*}+d_{12}\cos{\theta_2^*}),\,\,\,J_1=o\,(\mu),\\
&\,\,\,\,\quad\quad\theta_2=\theta_2^*+\frac{\mu}{\varkappa(J_{2}^0)^2}\cos{\theta_2^*},\,\,\,J_2=J_2^0-\frac{\mu}{\varkappa J_{2}^0}\sin{\theta_2^*};\\
&\widetilde{T}_2:\,\,\,\,\,
\theta_1=\beta_0+\varepsilon\alpha_0+\mu(d_{21}\sin{\theta_2^*}+d_{22}\cos{\theta_2^*}),\\
&\,\,\quad\quad J_1=J_2=J_2^0-\frac{\mu}{\varkappa J_{2}^0}\sin{(\beta_0+\varepsilon\alpha_0+\theta_2^*)},\\
&\,\,\quad\quad\theta_2=\theta_2^*-\mu(d_{21}\sin{\theta_2^*}+d_{22}\cos{\theta_2^*})+\frac{\mu}{\varkappa(J_{2}^0)^2}\cos{\theta_2^*},
\end{split}
\end{equation}
where $\varepsilon=\pm 1$ and constant coefficients $\,d_{ij}\,$ are expressed as
$$
d_{11}=\frac{\varkappa\sin{(\beta_0+\varepsilon\alpha_0)}+\varepsilon\sin{\alpha_0}[\cos{(\beta_0+\varepsilon\alpha_0)}-1]}{(J_2^0)^2(\varkappa^2+\sin^2{\alpha_0})},
$$
$$
d_{12}=\frac{\varkappa[1-\cos{(\beta_0+\varepsilon\alpha_0)}]+\varepsilon\sin{\alpha_0}\sin{(\beta_0+\varepsilon\alpha_0)}}{(J_2^0)^2(\varkappa^2+\sin^2{\alpha_0})},
$$
$$
d_{21}=\frac{\varkappa\sin{(\beta_0+\varepsilon\alpha_0)}-\varepsilon\sin{\alpha_0}[\cos{(\beta_0+\varepsilon\alpha_0)}-1]}{(J_2^0)^2(\varkappa^2+\sin^2{\alpha_0})},
$$
$$
d_{22}=\frac{\varkappa[1-\cos{(\beta_0+\varepsilon\alpha_0)}]-\varepsilon\sin{\alpha_0}\sin{(\beta_0+\varepsilon\alpha_0)}}{(J_2^0)^2(\varkappa^2+\sin^2{\alpha_0})}.
$$

The system (\ref{eq1}) solutions, relevant to the asymptotics of pendulum-like motions, come arbitrarily close to the limit cycles  $\,\widetilde{T}_1,\widetilde{T}_2$, furthermore the variable $\,\theta_1$ is in the range $[\beta_0+\varepsilon\alpha_0-\triangle_1,\beta_0+\varepsilon\alpha_0+\triangle_1],$
where
$$\,\,
\triangle_1=\mu\sqrt{d_{11}^2+d_{12}^2}\equiv\mu\sqrt{d_{21}^2+d_{22}^2}.
$$
After calculations we obtain
\begin{equation}
\label{dt}
\begin{split}\triangle_1=&\frac{\mu\,\sin{\alpha_0}\,\sin{\frac{1}{2}|\beta_0+\varepsilon\alpha_0|}}{\sqrt{\sin^2{\alpha_0}+16\sin^2{\frac{1}{2}(\alpha_0+\gamma_0)}\sin^2{\frac{1}{2}(\alpha_0-\gamma_0)}}\,h_0}=\\
=&\frac{\mu\left\vert r_1A_1\sqrt{A_2(A_3-A_2)}+\varepsilon
r_2A_2\sqrt{A_1(A_1-A_3)}\right\vert}{\sqrt{A_3(A_1-A_2)(4A_1A_3+4A_2A_3-3A_1A_2-4A_3^2)}\,h_0}.
\end{split}
\end{equation}
The same arguments may be used for investigation of the limit cycles, received from (\ref{tt}), with the replacement of $\beta_0+\varepsilon\alpha_0$ by $\beta_0+\varepsilon\alpha_0+2\pi.$

\subsection{Perturbed separatrices} As an example, let us consider one of possible variants. For this purpose, the following expressions
$$
\theta_1=\alpha_0+\beta_0+\delta{\theta_1},\,\,\,
J_1=J_1^*(t)+\delta{J_1},\,\,\,
\theta_2=\theta_2^*(t)+\delta{\theta_2},\,\,\,
J_2=J_2^0+\delta{J_2}
$$
should be assumed in the neighborhood of the separatrix (\ref{sol_2}), where the functions $\,J_1^*(t),$ $\theta_2^*(t)$ are defined by the formulas
(\ref{sol_2}). The linearized equations of the perturbed motion take the following form:
\begin{equation}
\label{eq2}
\begin{split}
&\delta\dot{\theta_1}=(2J_1^*-J_2^0)\sin{\alpha_0}\,\delta{\theta_1}+
\frac{\mu}{J_2^0}\left[\sin{(\alpha_0+\beta_0+\theta_2^*)}-\sin{\theta_2^*}\right],\\
&\delta\dot{J_1}=(J_2^0-2J_1^*)\sin{\alpha_0}\,\delta{J_1}+
J_1^*(J_2^0-J_1^*)\cos{\alpha_0}\,\delta{\theta_1}+\\
&\qquad\quad+J_1^*\sin{\alpha_0}\,\delta{J_2} -
\mu\,\frac{J_1^*}{J_2^0}\cos{(\alpha_0+\beta_0+\theta_2^*)},\\
&\delta\dot{\theta_2}=\varkappa\,\delta
J_2-J_1^*\sin{\alpha_0}\,\delta \theta_1+
\mu\,\frac{J_1^*}{(J_2^0)^2}\left[\sin{\theta_2^*}-\sin{(\alpha_0+\beta_0+\theta_2^*)}\right],\\
&\delta\dot{J_2}=-\mu\,\cos{\theta_2^*}+\mu\,\frac{J_1^*}{J_2^0}\left[\cos{\theta_2^*}-\cos{(\alpha_0+\beta_0+\theta_2^*)}\right].
 \end{split}
\end{equation}
The equations (\ref{eq2}) admit the first integral, depending on
$\,\delta\theta_1, \delta J_2:\,\,\,$
$$J=J_1^*(J_1^*-J_2^0)\sin{\alpha_0}\,\delta\theta_1+\varkappa J_2^0\delta
J_2+\mu\,\sin{\theta_2^*}+\mu\,\frac{J_1^*}{J_2^0}\left[\sin{(\alpha_0+\beta_0+\theta_2^*)}-\sin{\theta_2^*}\right].
$$
Differential equations (\ref{eq2}) are integrated in quadratures.
In particular, near the unperturbed separatrix the small deviation of $\,\theta_1$ from the initial value $\,\theta_1^0=\alpha_0+\beta_0$ we obtain from the first equation of (\ref{eq2}):
\begin{equation}
\label{dtheta}
\begin{split}
\delta\theta_1(t)=\,&\frac{J_1^0(J_2^0-J_1^0)}{J_1^*(J_2^0-J_1^*)}\,\delta\theta_1(t_0)+\frac{2\mu\sin{\frac{1}{2}(\alpha_0+\beta_0)}}{J_2^0J_1^*(J_2^0-J_1^*)}\times\\
&\times\int_{t_0}^tJ_1^*(J_2^0-J_1^*)\cos{(\theta_2^*+\frac{1}{2}(\alpha_0+\beta_0))}\,dt=\\
=\, &\frac{(e^{u(t-t_0)}+e^v)^2}{e^{u(t-t_0)}(e^v+1)^2}\,\delta\theta_1(t_0)+2\mu\sin{\frac{1}{2}(\alpha_0+\beta_0)}\,\frac{(e^{u(t-t_0)}+e^v)^2}{J_2^0\,e^{u(t-t_0)}}\times\\
&\times\int_0^{t-t_0}\hskip-4mm\frac{e^{u\tau}}{(e^{u\tau}+e^v)^2}\,\cos{(\varkappa
J_2^0\tau+\frac{1}{2}(\alpha_0+\beta_0)+\theta_2^0)}\,d\tau.
 \end{split}
\end{equation}

\section{Melnikov's integral}\label{sec4}
According to the notations~\cite{Koz80,Zig80}, we write the Hamiltonian (\ref{h0_new}) as sum $\,\mathcal{H}=\mathcal{H}_0+\mu\mathcal{H}_1.$  The distance between the stable and unstable separatrices of the system (\ref{eq1}) may be investigated analytically by means of the Melnikov's integral~\cite{Mel}
$$
I_M(\widetilde{\theta})=\int_{-\infty}^{\infty}\left\{
\mathcal{H}_0,\,\frac{\mathcal{H}_1}{\varkappa
J_2^0}\right\}(J_1^*,\theta_1^0,J_2^0,\theta_2^*+\widetilde{\theta})
d\,t,
$$
$$\left\{
\mathcal{H}_0,\,\frac{\mathcal{H}_1}{\varkappa
J_2^0}\right\}(J_1^*,\theta_1^0,J_2^0,\theta_2^*+\widetilde{\theta})=
\frac{1}{\varkappa J_2^0}\left.\left[\frac{\partial
\mathcal{H}_0}{\partial \theta_1}\frac{\partial
\mathcal{H}_1}{\partial J_1}-\frac{\partial \mathcal{H}_0}{\partial
J_1}\frac{\partial \mathcal{H}_1}{\partial
\theta_1}\right]\right\vert_{\,J_1^*,\theta_1^0,J_2^0,\theta_2^*+\widetilde{\theta}}\,,$$
evaluated along an unperturbed orbit (\ref{sol_2}), which connects hyperbolic periodic orbits.
The functions $\,J_1^*(t),\,\theta_1^0,$ $J_2^0,\,\theta_2^*(t)\,$ correspond to the doubly asymptotic solution (\ref{sol_2}). Owing to the transformations we find
$$
I_M(\widetilde{\theta})=\sigma_0\int_{-\infty}^{\infty}\frac{e^{u\tau-v}}{(e^{u\tau-v}+1)^2}\cos{(\varkappa
J_2^0\tau+\frac{1}{2}(\alpha_0+\beta_0)+\widetilde{\theta})}d\tau,
$$
in which
$\,\sigma_0=2\varkappa^{-1}\sin{\alpha_0}\sin{\frac{1}{2}(\alpha_0+\beta_0)}$
is nonzero constant. The integral $\,I_M(\widetilde{\theta})\,$ is a divergent improper integral, so we will be interested only in the evaluation of its principal value (the fast oscillating part is excluded). Using the linear substitution $\,\tau={(2\pi x+v)}/{u},$  we take into account the notations
$$
\nu=\frac{\pi
\varkappa}{\sin{\alpha_0}}>0,\,\,\,\,\,\delta=\frac{1}{2}(\alpha_0+\beta_0)+\frac{\varkappa\,
v}{\sin{\alpha_0}},
$$
The principal value $\,I_M(\widetilde{\theta})\,$ is reduced to Legendre improper integral, which is easily calculated:
\begin{equation}
\label{Im}
\begin{split}
\widehat{I}_M(\widetilde{\theta})&=\frac{\sigma_0}{u}\cos{(\delta+\widetilde{\theta})}\left[1-4\nu\int_{0}^{\infty}\frac{\sin{(2\nu
x)}}{e^{2\pi x}+1}\,d x\right]=\\
&=\frac{\sigma_0\nu}{u}\,\frac{\cos{(\delta+\widetilde{\theta})}}{\sinh{\nu}}=\frac{2\pi}{J_{2}^0}\,
\frac{\sin{\frac{1}{2}(\alpha_0+\beta_0)}}{\sin{\alpha_0}}\,\frac{\cos{(\delta+\widetilde{\theta})}}{\sinh{\nu}}.
\end{split}
\end{equation}
Therefore, for any values of parameters, restricted by the conditions (\ref{cond_P}), the integral $\,\widehat{I}_M(\widetilde{\theta}),$ considered as a function of the real argument  $\,\widetilde{\theta}$, has simple zeros only at the points $\,\widetilde{\theta}=\pm k\pi/2-\delta,\,$
$k=1,3,...\,.$ These values define two heteroclinic solutions of Hamiltonian
system (\ref{eq1}), asymptotically tending (if $\,t\to\pm\infty$) to two different periodic solutions $\,\widetilde{T}_1, \widetilde{T}_2$. The transversal intersection of the perturbed separatrices means that for the rigid body, that satisfies the conditions (\ref{cond_P}), chaotic motions always exist (at least, near the separatrices), if only the parameter $\,\mu\neq 0$ is sufficiently small in comparison to the energy constant $\,h_0.\,$

\begin{figure}\centering
\includegraphics[scale=0.5,clip]{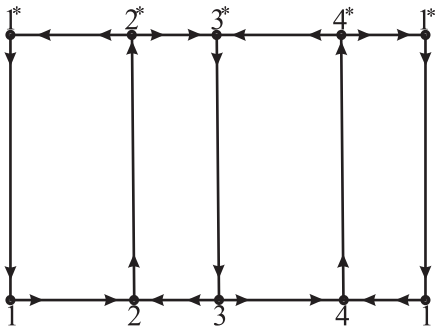}
\caption{Heteroclinic cycle at the plane $\mathbb{R}^2(\theta_1,J_1/J_2)$.}\label{p2}
\end{figure}

Following the methodology developed by Henri Poincar\'e, it is possible to prove,
that for fixed values of the parameters (\ref{cond_P}) and initial conditions, which correspond to $J_3=0,$ $h_0\gg\mu>0$, Hamiltonian equations (\ref{eq1}) admit a countable set of heteroclinic solutions. These solutions describe the doubly asymptotic pendulum-like motions of a rigid body. A heteroclinic cycle of the system (\ref{eq1}) is shown schematically at Fig.~\ref{p2}; it is composed of hyperbolic periodic orbits and heteroclinic orbits, which lie in intersection of perturbed separatrices.

An improper integral $\,\widehat{I}_M(\widetilde{\theta})\,$ is a special case of an integral evaluated by Ziglin~\cite{Zig80}. In order to check this statement it is possible to use the simplified formulas, obtained by Dovbysh in~\cite[p.~367]{Dov87}. The calculations, performed in this paper, confirm the well-known result of Kozlov~\cite[p.~104]{Koz80} about splitting of separatrices of asymmetric body, rotating about a fixed point in a weak field of gravity.

\section{Asymptotics of pendulum-like motions}\label{sec5}
The dynamical system (\ref{eq}) under the restrictions (\ref{cond_P}) has two hyperbolic periodic solutions $\,\,T_1\,$ and $\,T_2,$  which are represented by two closed nonintersecting curves in the space $\,\mathbb{R}^6({\bf M},{\boldsymbol \gamma})$.  Two asymptotic surfaces  $\,S_1^{\pm},S_2^{\pm}$ cross each of these curves. If $\,\mu=0\,$  these surfaces coincide pairwise, namely $\,S_1^{+}=S_2^{-},$ $S_2^{+}=S_1^{-}$. The Hess solution belongs to the double separatrix  $\,S_1^{+}=S_2^{-}$ which, as known from~\cite[\S~4.6]{Koz80}, is not split under perturbation (if $\; \mu\neq 0$). For the asymptotic motions of a rigid body in the case $\; h_0>\mu \;$ the following properties are fulfilled:
\begin{itemize}
    \item The invariant tori, which could isolate one of the following separatrices $\,S_1^{-},S_2^{-},S_1^{+},S_2^{+},$  are not existed;
    \item For fixed values of parameters of the dynamical system (\ref{eq}) at any energy level $h_0>\mu$  there is a countable set of heteroclinic solutions from $S_1^{-} \cap S_2^{+}$ $(S_2^{-} \cap S_1^{+})$ between Lyapunov periodic orbits  $\,T_1,\,T_2$;
    \item Any two trajectories on the separatrix $\,S_1^{-}\,(S_2^{+}) $  are separated by heteroclinic orbit, nearby trajectories diverge exponentially with time;
    \item Various types of homoclinic and heteroclinic orbits are characterized by a finite binary sequence $\,z(\widetilde\theta)$, coding the route along the unperturbed separatrices (\ref{sol_1})--(\ref{sol_22}).
   \end{itemize}
\begin{figure}[ht]\vskip-3mm\centerline{
\includegraphics[scale=1.20,clip,viewport=5 5 52 105]{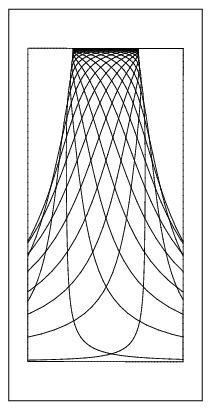}\hskip1mm
\includegraphics[scale=1.20,clip,viewport=5 5 52 105]{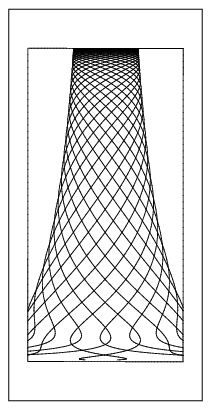}\hskip1mm
\includegraphics[scale=1.20,clip,viewport=5 5 52 105]{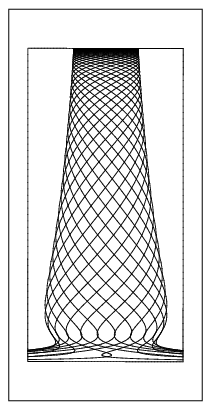}\hskip1mm
\includegraphics[scale=1.20,clip,viewport=5 5 52 105]{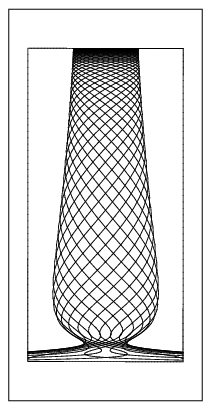}\hskip1mm
\includegraphics[scale=1.20,clip,viewport=5 5 52 105]{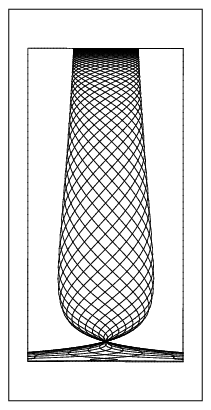}\hskip1mm
\includegraphics[scale=1.20,clip,viewport=5 5 52 105]{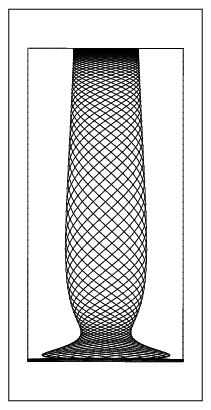}\hskip1mm
\includegraphics[scale=1.20,clip,viewport=5 5 52 105]{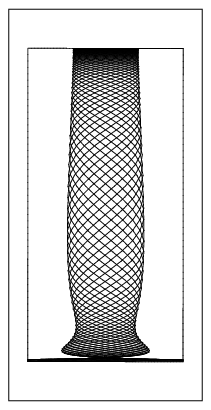}
} \centerline{
\includegraphics[scale=1.20,clip,viewport=5 5 52 105]{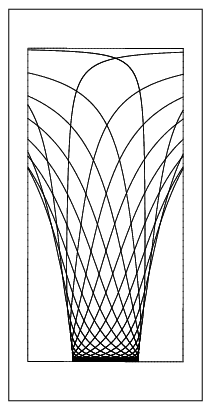}\hskip1mm
\includegraphics[scale=1.20,clip,viewport=5 5 52 105]{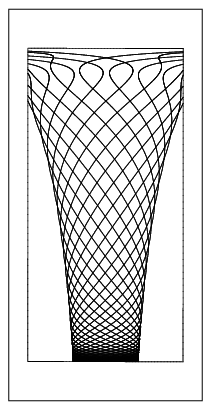}\hskip1mm
\includegraphics[scale=1.20,clip,viewport=5 5 52 105]{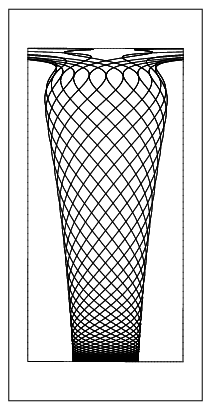}\hskip1mm
\includegraphics[scale=1.20,clip,viewport=5 5 52 105]{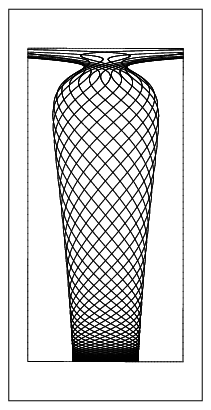}\hskip1mm
\includegraphics[scale=1.20,clip,viewport=5 5 52 105]{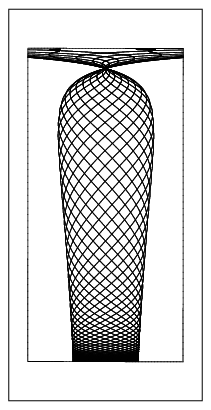}\hskip1mm
\includegraphics[scale=1.20,clip,viewport=5 5 52 105]{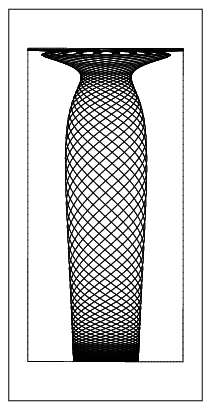}\hskip1mm
\includegraphics[scale=1.20,clip,viewport=5 5 52 105]{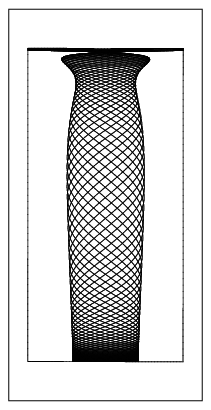}
} \centerline{\hskip0mm{\footnotesize\it
a}\hskip20mm{\footnotesize\it b}\hskip20mm{\footnotesize\it
c}\hskip20mm{\footnotesize\it d}\hskip20mm{\footnotesize\it
e}\hskip20mm{\footnotesize\it f}\hskip20mm{\footnotesize\it g}}
\caption{Asymptotic surfaces $S_1^{-}, S_2^{+}$.}\label{p3}
\end{figure}

Projections of phase trajectories of the perturbed system (\ref{eq1}) in $\mathbb{R}^2(\theta_1,J_1/J_2)$  are shown at Fig.~\ref{p3}. The calculations were carried out for the following values of the parameters:
$$\mu=0.01,\,\,\beta_0=\gamma_0,\,\,
\theta_1^0=\alpha_0+\beta_0,\,\,J_2^0=2/\sqrt{A_3\varkappa},\,\,\theta_2^0\in(0,2\pi);$$
the values of $\alpha_0,\,\gamma_0$ can be found in Table~\ref{t1}.
\begin{table}[ht]
\centering
 \begin{tabular}{|c||c|c|c|c|c|c|c|}\hline
 &\textit{a}&\textit{b}&\textit{c}&\textit{d}&\textit{e}&\textit{f}&\textit{g}\\
\hline
\hline $\alpha_0$&2.91891&2.23654&1.23096&1.61146&1.97724&0.79020&0.56207\\
\hline $\gamma_0$&2.46192&1.95519&0.92730&1.34948&1.77215&0.57351&0.40272\\
\hline
\end{tabular}
\vskip5mm
\caption{The values $\,\alpha_0,\gamma_0\,$ for Fig.~\ref{p3}.}\label{t1}
\end{table}
At the lower part of Fig.~\ref{p3} the surfaces $S_1^{-},$ formed by the solutions of the system (\ref{eq1}), are shown, which in case $t\to-\infty$ are asymptotically approached to the periodic solution $\,\widetilde{T}_1.$ In the upper part of Fig.~\ref{p3}  the surfaces $S_2^{+},$ formed by the solutions of the system (\ref{eq1}), are shown, which in case $t\to+\infty$ are asymptotically approached to the periodic solution $\,\widetilde{T}_2.$  The set $S_0=(S_1^{-}\cap S_2^{+})\cup(S_2^{-}\cap S_1^{+})$ consists of heteroclinic solutions of the system (\ref{eq1}). Homoclinic solutions of the system (\ref{eq1}) belong to the set $S^0=(S_1^{-}\cap
S_1^{+})\cup(S_2^{-}\cap S_2^{+})$.

\begin{figure}[ht]
\centering
\includegraphics[scale=1.8,clip,viewport=20 5 182
80]{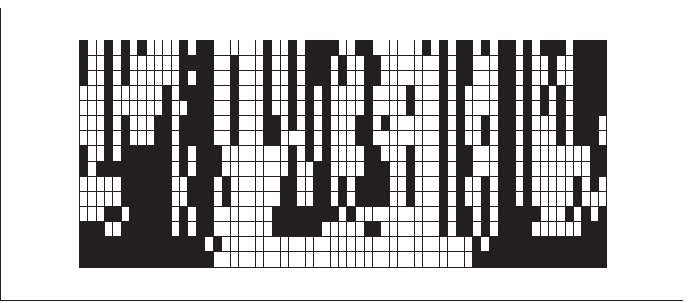}\hskip2mm
\includegraphics[scale=1.8,clip,viewport=5 5 82 80]{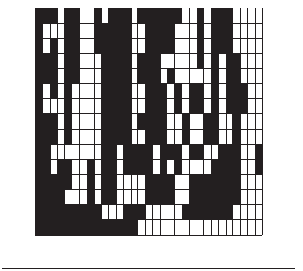}
\caption{The sequence $z(\widetilde\theta_k)$ for the orbits on $S_1^{-}.$}\label{p4}
\end{figure}
At Fig.~\ref{p4} one can see the possible routes of trajectories on $\,S_1^{-}\, $ in the vicinity of the unperturbed separatrices (\ref{sol_1})--(\ref{sol_22}). Using the points $\widetilde\theta_k=k/10$, where $k=\overline{0,62}$, we define a partition of the interval $[0, 2\pi)$, in which Melnikov's function (\ref{Im}) argument $\,\widetilde\theta\,$  is changed. Let us consider the set of solutions for the system (\ref{eq1}),  subject to the initial conditions $\theta_1^0=\alpha_0+\beta_0,\,\,J_1^0\approx 0,$
$\,\,\theta_2^0=\widetilde\theta_k,\,\,J_2^0=\textrm{const},$ in a small neighborhood of periodic solution $\,\widetilde{T}_1.$ Near the solution $\,\widetilde{T}_2,$ according to (\ref{dtheta}), the phase orbit leaves the neighborhood of the separatrix (\ref{sol_2}) and, furthermore, moves along the perimeter of the left or right rectangle at Fig.~\ref{p2}.
Let $\; z_i(\widetilde\theta_k)=0,$ if the trajectory is displaced on the left from the incoming separatrix, and  $\;
z_i(\widetilde\theta_k)=1,$ if the trajectory is displaced on the right from the incoming separatrix, Fig.~\ref{p2}; $i$  is the index of the sequence of hyperbolic periodic solutions, in the neighborhood of which the solution of the system (\ref{eq1}) under consideration has been running across. The columns at Fig.~\ref{p4} are composed of the first members of the sequences $\; z(\widetilde\theta_k)$, where black and white cells coincide with 0 and 1 of $\; z_i(\widetilde\theta_k),
i\in\mathbb{Z}$. The calculations have been rendered for $i=\overline{1,15}$ and the following values of the parameters: $\,A_1=2,\, A_2=1.0,\, A_3=1.5,$
${\bf r}=(\sqrt{2}/2,\sqrt{2}/2,0),\mu=0.001, h_0=1.3342$. The bottom row of the array at Fig.~\ref{p4} corresponds to $i=1$, the top row is that of $i=15.$ Black and white cells at the bottom row, Fig.~\ref{p4}, correspond to different signs of Melnikov's function (\ref{Im}). Right array of Fig.~\ref{p4} is a magnification of some part of the left array: using the following points $\widetilde\theta_k=3\pi/40+k/100$, where $k=\overline{0,31}$, the partition of the interval $[3\pi/40, 7\pi/40]$  is defined. Multiple zooming of a partition step has no effect on the general qualitative properties of the phase orbits under investigation.

\section{Body motion in a fixed basis} \label{sec6}
We will study the qualitative properties of body motion in a fixed basis. With zero constant of the momentum integral $\,( J_3=0 )\,$  the angular momentum $\,\bf{M}\,$ during all time of motion, lies in a horizontal plane $\,O\eta\zeta.$  The direction of vector $\,\bf{M}\,$  in a plane $\,O\eta\zeta\,$  and its modulus $\,\vert\bf{M}\vert\,$ are characterized by variables $\theta_3, J_2,$ therefore the time evolution of the angular momentum is described by the functions $\theta_3(t), J_2(t).$ It is well known that the pendulum motions, which are described by (\ref{sol_3}), must satisfy the additional restriction $\,\theta_3=\textrm{const},$ i.e. the direction of $\,\bf{M}\,$ is preserved in a fixed basis, but the modulus $\,\vert\bf{M}\vert$ is changed.
For the perturbed separatrices, described by the linearized equations of type (\ref{eq2}), it is possible to find the increment of angle $\,\theta_3\,$ with increasing time to infinity:
\begin{equation}\label{tri1}
\begin{split}
&1)\,\,\theta_1^0=\alpha_0-\beta_0,\,\,\,\triangle^{(1)}\theta_3\approx\frac{2\mu\pi}{(J_2^0)^2}\frac{\sin{\frac{1}{2}(\beta_0-\alpha_0)}}{\sin{\alpha_0}}\approx\frac{\mu\pi\varkappa}{h_0}\frac{\sin{\frac{1}{2}(\beta_0-\alpha_0)}}{\sin{\alpha_0}},\\
&2)\,\,\theta_1^0=\alpha_0+\beta_0,\,\,\,\triangle^{(2)}\theta_3\approx\frac{2\mu\pi}{(J_2^0)^2}\frac{\sin{\frac{1}{2}(\beta_0+\alpha_0)}}{\sin{\alpha_0}}\approx\frac{\mu\pi\varkappa}{h_0}\frac{\sin{\frac{1}{2}(\beta_0+\alpha_0)}}{\sin{\alpha_0}},\\
&3)\,\,\theta_1^0=\alpha_0-\beta_0+2\pi,\,\,\,\triangle^{(3)}\theta_3=-\triangle^{(1)}\theta_3,\\
&4)\,\,\theta_1^0=\alpha_0+\beta_0+2\pi,\,\,\,\triangle^{(4)}\theta_3=-\triangle^{(2)}\theta_3.
\end{split}
\end{equation}

In particular, for Hess case $\,(\alpha_0=\beta_0)$ in paper~\cite{KGK} the following expression has been obtained
\begin{equation}
\nonumber
\triangle^{(2)}\theta_3\approx\frac{2\mu\pi}{(J_2^0)^2}\approx\frac{\mu\pi
A_1A_2}{A_3(A_1-A_2)h_0}\,.
\end{equation}
It should be noted, that this value can be large enough, if the inertia ellipsoid is slightly different from the sphere.

For Grioli case $\,(\gamma_0=\beta_0)\,$ from (\ref{tri1}), including (\ref{phi}) we obtain
$$
\triangle^{(1)}\theta_3\approx-\frac{\mu\pi}{4\,h_0}\frac{\sin{\alpha_0}}{\sin{\frac{1}{2}(\alpha_0+\gamma_0)}}\in\left(\frac{\mu\pi}{h_0}\frac{(1-\sqrt{2})}{\sqrt{2}},0\right),
$$
$$
\triangle^{(2)}\theta_3\approx\frac{\mu\pi}{4\,h_0}\frac{\sin{\alpha_0}}{\sin{\frac{1}{2}(\alpha_0-\gamma_0)}}\in\left(0,\infty\right).
$$

The angle $\,\theta_3\,$ is the angle of precession of angular momentum about the vertical axis. In space of rigid body parameters we can distinguish subsets with different angle $\,\theta_3\,$  behavior in the neighborhood of perturbed separatrices:
\begin{equation}\label{trieq}
\begin{split}
&{\it{i}})\,\,\,\,\,\pi=\beta_0>\alpha_0>0,\,\,\,\triangle^{(2)}\theta_3=\triangle^{(1)}\theta_3>0>\triangle^{(3)}\theta_3=\triangle^{(4)}\theta_3;\\
&{\it{ii}})\,\,\,\pi>\beta_0>\alpha_0>0,\,\,\,\triangle^{(2)}\theta_3>\triangle^{(1)}\theta_3>0>\triangle^{(3)}\theta_3>\triangle^{(4)}\theta_3;\\
&{\it{iii}})\,\,\pi>\beta_0=\alpha_0>0,\,\,\,\triangle^{(2)}\theta_3>\triangle^{(1)}\theta_3=0=\triangle^{(3)}\theta_3>\triangle^{(4)}\theta_3;\\
&{\it{iv}})\,\,\,\pi>\alpha_0>\beta_0>0,\,\,\,\triangle^{(2)}\theta_3>\triangle^{(3)}\theta_3>0>\triangle^{(1)}\theta_3>\triangle^{(4)}\theta_3;\\
&{\it{v}})\,\,\,\,\,\pi>\alpha_0>\beta_0=0,\,\,\,\triangle^{(2)}\theta_3=\triangle^{(3)}\theta_3>0>\triangle^{(1)}\theta_3=\triangle^{(4)}\theta_3.
\end{split}
\end{equation}
Thus, for small values of $\,\mu\,$  the perturbed motion orbit returns to a small neighborhood of separatrices (\ref{sol_1})--(\ref{sol_22}), in this case the angle $\,\theta_3\,$ gets the increment $\,\triangle^{(i)}\theta_3\,$ in agreement with formulas (\ref{tri1}). At the other areas of motion near boundaries  $J_1=0,$ $J_2-J_1=0$
we have $\,\triangle\theta_3=o\,(\mu).$ The distinctive feature of heteroclinic and homoclinic solutions is the existence of finite limits
$$
\underset{t\to-\infty}{\lim}{\theta_3(t)}=\theta_3^{-},\,\,\,\,\,\underset{t\to+\infty}{\lim}{\theta_3(t)}=\theta_3^{+}.
$$
In general case, doubly asymptotic solutions orbits are considered to be the heteroclinic in a fixed space, but it is possible to select the dynamical system parameters in such a way, that the following difference $\,(\theta_3^{+}-\theta_3^{-})\neq0\,$ could be any value, e.g., a rational multiple of $\,2\pi.$

As a geometric interpretation of a rigid body motion let us consider Poinsot kinematic representation by the rolling motion of the body's ellipsoid of inertia on a fixed plane in space. Let us $\,\mu=0.$ The inertia ellipsoid for the fixed point $O$ is defined by the equation
\begin{equation}
\label{el}A_1x^2+A_2y^2+A_3z^2=1.
\end{equation}
We denote by $\rho$ value the radius-vector of this ellipsoid, directed on the instantaneous axis of rotation, and draw the tangent plane $\Pi$  through the end of the indicated radius-vector. Then the inertia ellipsoid (\ref{el}) during the motion rolls without sliding on one of its tangent planes, this plane $\Pi$ is orthogonal to the angular momentum ${\bf M}$ and remains fixed in the space.
The angular velocity ${\boldsymbol
\omega}$ of a moving body is directed along the radius-vector of a point of contact, and its value is proportional to $\rho.$   Louis Poinsot developed a visualization to the motion of the endpoint of the angular velocity vector. The path traced out on the inertia ellipsoid by the angular velocity vector ${\boldsymbol
\omega}$ is called the polhode; the corresponding curve on the invariable plane is called the herpolhode. Geometric properties of polhodes and herpolhodes essentially depend on the values of moments of inertia and initial conditions of motion.

The polhode is a closed curve encircling the major or minor axis of inertia ellipsoid. The boundary curve, which splits two families of polhodes, consists of two intersecting ellipses, passing through  the intermediate axis of inertia. In this case $\vert{\bf M}\vert^2=2A_3h_0,$ then the polhode is called the separatrix. Herpolhode, corresponding to polhode-separatrix, is a symmetric spiral in the fixed plane $\Pi,$ which runs around the $Q$ -- an intersection point of the momentum ${\bf M}$ with the plane $\Pi.$ The length of this spiral is finite, and equals to the corresponding polhode arc.

\begin{figure}[ht]
\centerline{
\includegraphics[scale=1.0,clip,viewport=3 2 204
100]{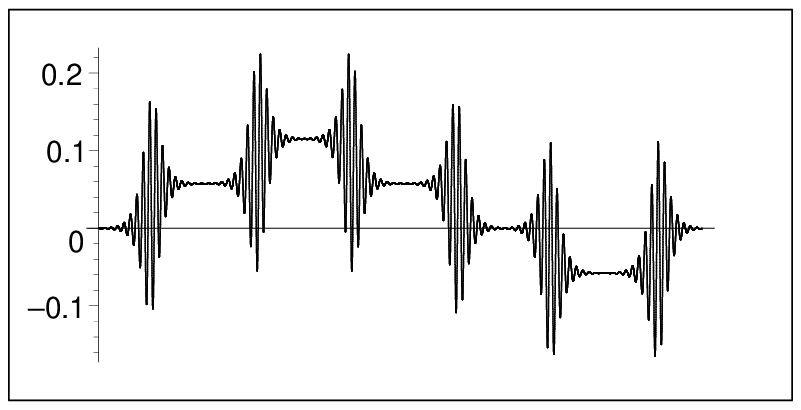}\hskip10mm
\includegraphics[scale=0.5,clip,viewport=20 11 245
195]{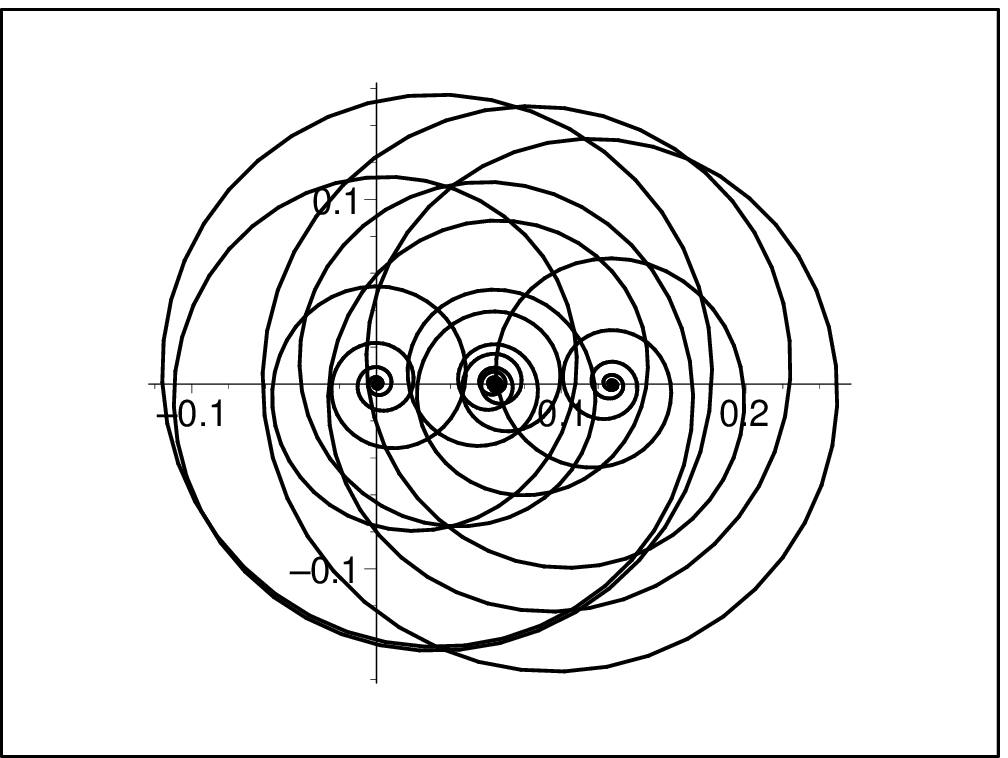}} \centerline{\hskip30mm{\footnotesize {\it
a)\,\,$\alpha(t)$ for $\theta_2^0=1.02$}}\hskip25mm{\footnotesize
{\it b)\,\,$(\omega_\zeta(t),\omega_\xi(t))$ for
$\theta_2^0=1.02$}}} \centerline{
\includegraphics[scale=1.0,clip,viewport=3 2 204
100]{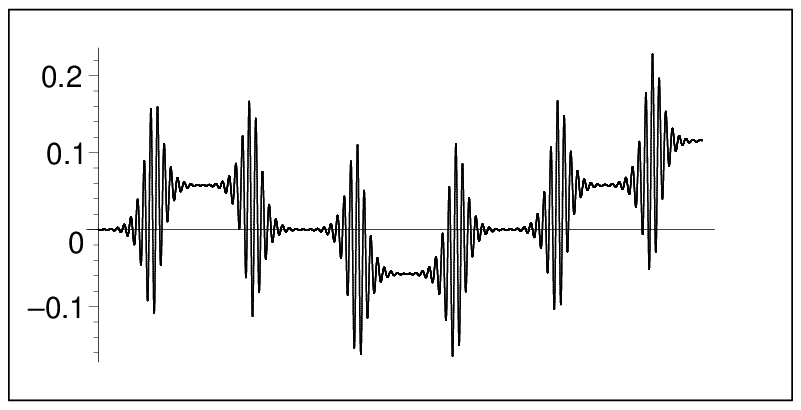}\hskip10mm
\includegraphics[scale=0.5,clip,viewport=20 11 245
195]{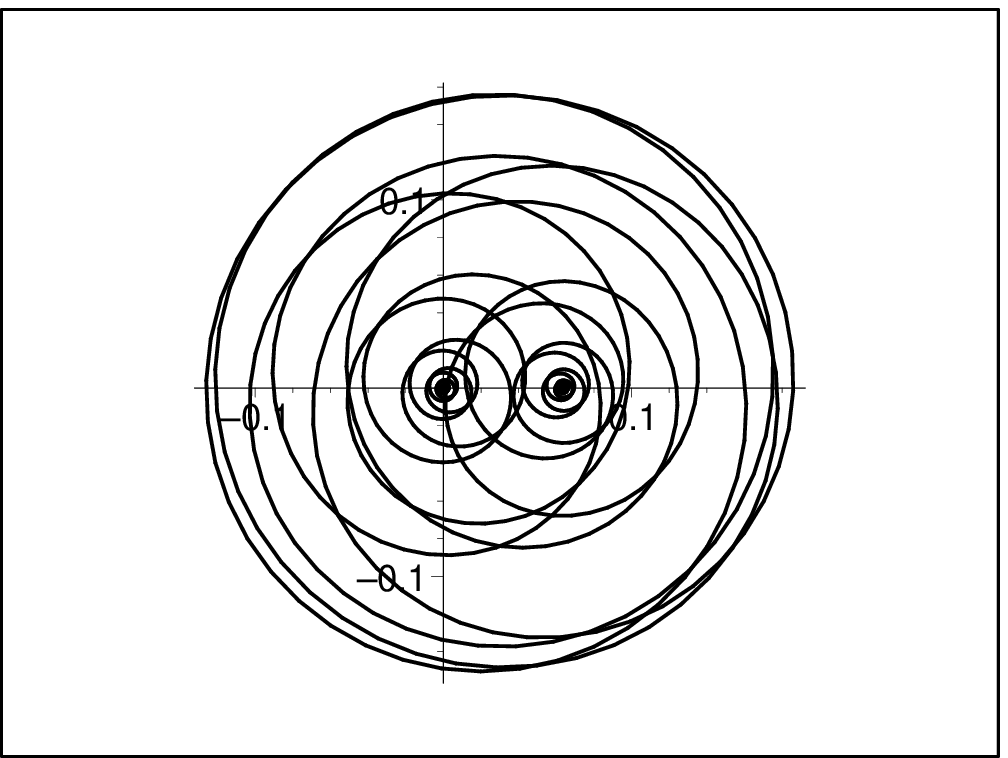}} \centerline{\hskip30mm{\footnotesize {\it
c)\,\,$\alpha(t)$
 for $\theta_2^0=1.83$}}\hskip25mm{\footnotesize {\it d)\,\,$(\omega_\zeta(t),\omega_\xi(t))$ for
 $\theta_2^0=1.83$}}}
\caption{Perturbed herpolhodes in a neighborhood of pendulum-like motions.}\label{p5}
\end{figure}

For $\mu\neq 0$ we write the equations of perturbed Poinsot herpolhode. For this purpose it is necessary to find the components of angular velocity vector $\,{\boldsymbol \omega}=\omega_{\xi}{\bf i}_1+\omega_{\eta}{\bf
i}_2+\omega_{\zeta}{\bf i}_3\,$  in a fixed orthonormal basis $\,({\bf i}_1,{\bf i}_2,{\bf i}_3).$

Assuming, that $\,\omega_{\eta}=\omega_{\rho}\cos{\alpha},\,
\omega_{\zeta}=\omega_{\rho}\sin{\alpha},$ we can use the Kharlamov's~\cite{PVKH65} kinematic equations describing the fixed hodograph of an angular velocity:
\begin{equation}\label{eqPV}
\omega_{\xi}\equiv{\boldsymbol \omega}\cdot{\boldsymbol \gamma},\,\,\,\,\,
\omega_{\rho}^2\equiv\vert{\boldsymbol \omega}\vert^2-({\boldsymbol \omega}\cdot{\boldsymbol \gamma})^2,\,\,\,\,\,
\omega_{\rho}^2\,\frac{d\alpha}{d
t}=({\boldsymbol \gamma}\times{\boldsymbol \omega})\cdot\frac{d{\boldsymbol \omega}}{d t}.
\end{equation}
For the pendulum motions (\ref{5:3}) let us assume, that $\omega_{\xi}=\omega_{\zeta}=0.$  Under a small perturbation of the pendulum rotations  the endpoint of the angular velocity vector traces out the curve $(\omega_\xi(t),\omega_\eta(t),\omega_\zeta(t))$ in the fixed space, its projection on the plane $\mathbb{R}^2(\omega_\zeta,\omega_\xi)$  is a spiral: a planar curve that winds around the origin of coordinates. Near the separatrices (\ref{sol_1})--(\ref{sol_22}) the angle $\alpha(t)$ receives the nonzero increment, defined by formulas (\ref{tri1}). Indeed, under the restriction $J_3=0$ the angles $\theta_3,\,\alpha\,$ are connected by the following relation
\begin{equation}\label{th_al}\theta_3(t)=\alpha(t)-\textrm{arctan}
\left(\frac{\bf{M}\cdot({\boldsymbol \omega}\times{\boldsymbol \gamma})}{\bf{M}\cdot{\boldsymbol \omega}}\right).
\end{equation}

At Fig.~\ref{p5} for two values of $\theta_2^0$  the graphs of the function $\alpha(t)$  and the curves $(\omega_\zeta(t),\omega_\xi(t))$ are presented. These calculations were carried out for the following values of the parameters:
$$A=(2,1.5,1.8),\, {\bf r}=(0,1,0),\, \mu=0.01,\, h_0=1.1111.$$
In this case the following conditions (\ref{trieq}),{\it i}):
$\triangle^{(1)}\theta_3=\triangle^{(2)}\theta_3=-\triangle^{(3)}\theta_3=-\triangle^{(4)}\theta_3\approx 0.057715$ have been fulfilled. The motion along separatrices (\ref{sol_1})--(\ref{sol_22}), as it follows from the results of the numerical integration of the system (\ref{eq1}), may be described by the ordered sequence of the heteroclinic cycle tops at Fig.~\ref{p2}:
\begin{equation}\nonumber
\begin{split}
\theta_2^0=1.02:&\quad \{2-2^*-1^*-1-4-4^*-3^*-3-4-4^*-1^*-1\};\\
\theta_2^0=1.83:&\quad \{2-2^*-3^*-3-4-4^*-1^*-1-2-2^*-1^*-1\}.
\end{split}
\end{equation}
The change in the angle $\alpha(t)$ depends on the route of motion along unperturbed separatrices (see Fig.~\ref{p5},{\it a,c}). At Fig.~\ref{p5},{\it b,d} the curves $(\omega_\zeta(t),\omega_\xi(t))$ are shown, which coincide with the different routes of motion from the top 2 to the top $4^*$ of heteroclinic cycle.

\end{document}